\documentclass[acmsmall,screen]{acmart}

\AtBeginDocument{%
  \providecommand\BibTeX{{%
    \normalfont B\kern-0.5em{\scshape i\kern-0.25em b}\kern-0.8em\TeX}}}

\setcopyright{acmcopyright}
\copyrightyear{2025}
\acmYear{2025}
\acmDOI{10.1145/1122445.1122456}

\acmJournal{JACM}
\acmVolume{1}
\acmNumber{1}
\acmArticle{1}
\acmMonth{1}

\usepackage{color}
\usepackage{amssymb}
\usepackage{booktabs}
\usepackage{amsmath}
\usepackage{subfigure}
\usepackage{svg}

\usepackage{url}
\usepackage{algorithm}
\usepackage{algpseudocode}

\usepackage{threeparttable}     % 涉及到表格的列宽
\usepackage{multirow}
\usepackage{subfigure}
\usepackage{float}
\usepackage{makecell}
\usepackage{tabularx} 
\usepackage{graphicx}

\usepackage{caption}
\usepackage{subcaption}
\usepackage{color}
\usepackage{xspace}
\usepackage{tcolorbox}
\usepackage{tikz}
\newcommand*{\circled}[1]{\lower.7ex\hbox{\tikz\draw (0pt, 0pt)%
    circle (.45em) node {\makebox[0.6em][c]{\small #1}};}}

\begin{document}

\title{A Novel Interactive-Guided Differential Testing Approach for FPGA Simulation Debugger Tools}

%% Of note is the shared affiliation of the first two authors, and the　"authornote" and "authornotemark" commands　used to denote shared contribution to the research.

\author{Shikai Guo}
\email{shikai.guo@dlmu.edu.cn}
\affiliation{%
  \institution{Dalian Maritime University}
   \country{China}
}

\author{Xiaoyu Wang}
\email{xiaoyuwang@dlmu.edu.cn}
\affiliation{%
  \institution{Dalian Maritime University}
   \country{China}
}

\author{Xiaochen Li}
\email{xiaochen.li@dlut.edu.cn}
\affiliation{%
  \institution{Dalian University of Technology}
   \country{China}
}

\author{Zhihao Xu}
\email{Hosea.xu@seu.edu.cn}
\affiliation{%
  \institution{SouthEast University}
  \streetaddress{2 SouthEast Universty Rd}
  \city{Nanjing}
  \state{Jiangsu}
  \country{China}
}

\author{He Jiang}
\email{jianghe@dlut.edu.cn}
\affiliation{%
  \institution{Dalian University of Technology}
   \country{China}
}

%%
%% By default, the full list of authors will be used in the page
%% headers. Often, this list is too long, and will overlap
%% other information printed in the page headers. This command allows
%% the author to define a more concise list
%% of authors' names for this purpose.
\renewcommand{\shortauthors}{Guo et al.}

%% By default, the full list of authors will be used in the page
%% headers. Often, this list is too long, and will overlap
%% other information printed in the page headers. This command allows
%% the author to define a more concise list
%% of authors' names for this purpose.
% \shortauthors{Trovato and Tobin, et al.}

%% Abstract.
\begin{abstract}

Field-Programmable Gate Array (FPGA) development tool chains are widely used in FPGA design, simulation, and verification in critical areas like communications, automotive electronics, and aerospace. 
Commercial FPGA tool chains such as Xilinx’ Vivado aids developers in swiftly identifying and rectifying bugs and issues in FPGA designs through a robust built-in debugger, ensuring the correctness and development efficiency of the FPGA design.
Hardening such FPGA chip debugger tools by testing is crucial since engineers might misinterpret code and introduce incorrect fixes, leading to security risks. 
However, FPGA chip debugger tools are challenging to test as they require assessing both RTL designs and a series of debugging actions, including setting breakpoints and stepping through the code.
To address this issue, we propose a interactive differential testing approach called DB-Hunter to detect bugs in Vivado's FPGA chip debugger tools.
Specifically, DB-Hunter consists of three components: RTL design transformation component, debug action transformation component, and interactive differential testing component.
By performing RTL design and debug action transformations, DB-Hunter generates diverse and complex RTL designs and debug actions, to thoroughly test the Vivado debugger using interactive differential testing to detect bugs. 
In three months, DB-Hunter reported 18 issues, including 10 confirmed as bugs by Xilinx Support, 6 bugs had been fixed in last version.

\end{abstract}

% \ccsdesc[500]{Software and its engineering~Collaboration in software development}
% \ccsdesc[500]{Computing methodologies~Neural networks}
% \ccsdesc[500]{Computing methodologies~Natural language processing}
%% Keywords.

\maketitle

\section{Introduction}
Field-Programmable Gate Array (FPGA) Engineers rely heavily on complex simulation verification tools to design FPGA chips (e.g., MicroBlaze and ARM Cortex-M) \cite{hartley2013predictive} to utilize in crucial domains such as communication, automotive electronics, and aerospace \cite{bai2023soft,sayeed2019eseiz,paikrao2023consumer}. 
Such FPGA chip debugger tools could help developers in swiftly identifying and rectifying bugs in FPGA designs through a robust built-in debugger, which assists engineers in the rapid and precise development and debugging of FPGA projects.
Commercial FPGA chip debugger tools, such as the built-in debugger in Xilinx's Vivado, enable developers to directly trace the runtime behavior of FPGA designs. 
It allows engineers to pause execution at points of interest using breakpoints, inspect the intermediate program state (such as the values of local variables), and closely follow the control flow through single-stepping.

When the debugger of the FPGA chip debugger tools have a bug, it can cause engineers to make incorrect changes to the program, which leads to the potential risk. In addition, as the scale of FPGA usage increases and the complexity of logic circuits customizable, hardening such FPGA chip debugger tools (such as Vivado debugger) by testing is crucial since engineers might misinterpret code and introduce incorrect fixes, leading to development bugs or inject subtle unexpected behaviors \cite{drimer2009security,moradi2011vulnerability}.

\begin{figure}[!t]
  \centering
  \includegraphics[width=0.75\linewidth]{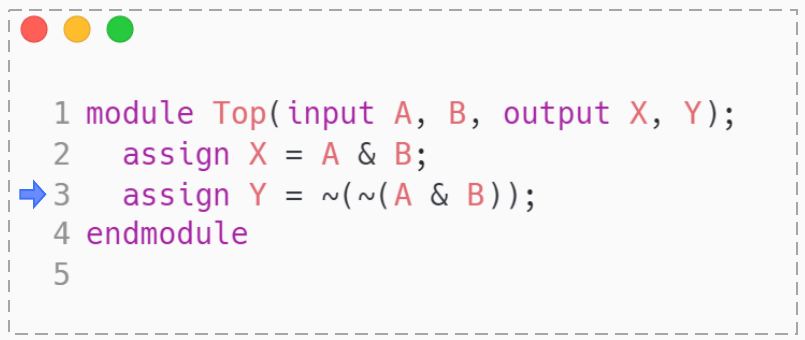}
  \begin{center}
      (a) Initial test run
  \end{center}
  \includegraphics[ width=0.75\linewidth]{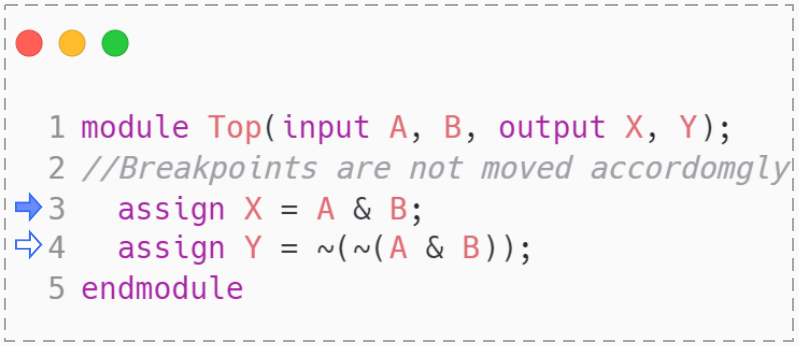}
  \begin{center}
      (b) Follow-up test run
  \end{center}
  \caption{A bug about breakpoint misplacement}\label{fig.1}
\end{figure}

An example of this type of bug is shown in \figurename~\ref{fig.1} (\includegraphics[width=0.2cm]{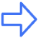}: expected breakpoint location,  \includegraphics[width=0.2cm]{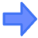}: actual breakpoint location)\footnote{\url{https://support.xilinx.com/s/feed/0D54U00007QysW8SAJ?language=en_US}}.
We anticipate placing a breakpoint at a certain location, but due to the Vivado debugger bug, the breakpoints do not move along with the program after inserting blank lines or comments. 
This behavior may lead engineers to mistakenly believe that lines with breakpoints have not been executed at all, resulting in incorrectly changes to the FPGA design code.
Specifically, \figurename~\ref{fig.1}(a) shows a breakpoint inserted at line 3. 
Subsequently, a blank line is added after line 1, causing the code below line 1 to shift down by one line, and the breakpoint should move down with it.
However, during actual execution, we observed that the breakpoint at line 3 did not move along with the original code at line 3, as shown in \figurename~\ref{fig.1}(b).
Since the Vivado debugger inaccurately describes the actual execution of the FPGA design code, engineers may make changes to the program for worse (e.g., error fixes).

One effective approach for detecting bugs in debuggers is through automated or random testing. 
In this process, test inputs are randomly generated, and a test oracle~\cite{barr2014oracle} is employed to assess the success or failure of the tests.
Notably, automation technology tailored for the Chromium browser's Javascript debugger has recently been introduced \cite{lehmann2018feedback}. 
The technique randomly generates debug actions, such as setting breakpoints, resuming execution, or single-stepping through code, based on syntax. 
Then, these actions are applied to a specified program slated for debugging. 
Finally, a differential testing is used to compare the debug traces output by the debugger. 
However, the differing syntactic structures and statement execution sequences between JavaScript and Verilog languages render the conversion approaches applicable to debugging JavaScript debuggers ineffective for Verilog debuggers. 
Therefore, a specialized conversion approach specifically tailored for the Verilog language is required.

However, due to the interactive nature of the debugger, debug actions persist throughout the entire debugging session. 
The results in debug action transformations sometimes rely on the current state of the debugger and necessitate interactive transformations of debug actions during the testing process. 
How to build equivalent debugging operations suitable for debugger testing during debugging is a significant challenge.

% Additionally, the debugger accepts both transforming RTL designs and debugging actions (e.g., breakpoints and steps) as input. 
% This necessitates addressing two distinct transformation approaches: RTL design transformation and debug action transformation. 
% As debugging actions are linked to code locations (e.g., the line used to set a breakpoint), consistency between the two transformations is crucial. 
% For example, if during the RTL design transformation process, the deletion of code results in a change in line numbers, the associated debug actions should be updated accordingly. 
% Thus to ensure consistency between the two transformation approaches is another challenge.

To overcome these challenges, we propose DB-Hunter, a approach for bug detection in Vivado debugger tool using interactive differential testing. 
DB-Hunter consists of three main components: RTL design transformation component, debug action transformation component, and interactive differential testing component. 
By transforming RTL designs and debugging actions separately, DB-Hunter generates a more varied and complex set of RTL design variants.
The increased complexity introduced by these RTL design variants may lead to the triggering of more bugs during the debugging process in the Vivado debugger.
Since these RTL design variants do not affect the functionality of the original RTL design and the Vivado debugger, they are considered equivalent. 
Finally, using the variant files generated post-program and debug action transformations, we rerun Vivado and record debugging traces. 
If DB-Hunter detects inconsistencies in the output of RTL design and debug actions, it may indicate the presence of bugs in the debugger.

In addition, DB-Hunter is iterative, wherein the RTL design variants generated by one transformation mode can serve as input for other types of transformations. 
We can thus apply the transformations a second time on the follow-up test run to obtain yet another test input. 
With these new inputs, we can apply program and action transformations again, or even continue indefinitely. 
If the comparison on the debugging traces is successful, iterative test continues by transforming the inputs of the follow-up RTL design once more with a randomly selected transformation. 
This process continues until the RTL design fails (and a potential bug is found) or until we reach a maximum number of iteration rounds.

Experiments shows that DB-Hunter excels in detecting bugs in the Vivado debugger. In three months, DB-Hunter has detected 18 bugs within Vivado, where 10 have been officially confirmed by Xilinx\footnote{\url{https://support.xilinx.com/s/topic/0TO2E000000YKY4WAO/simulation-verification?language=en_US}} as new bug issues, and 6 has been fixed. 
In addition, we also verified the ability of the DB-Hunter transformation component and different iteration numbers to detect Vivado debugger bugs.

This article presents the following contributions:
\begin{itemize}
\item To the best of our knowledge, we are the first to undertake debugger testing within Vivado. We propose a custom program transformation and debug action transformation approaches specifically designed for Verilog language to solve the challenges related to debugger testing in Vivado. 
\item We validated DB-Hunter in Vivado 2022.1 and Vivado 2023.1, and detected 18 previously undiscovered bugs, with 10 confirmed by the Xilinx official team as new issues, and 6 had been fixed.
\item We release the source code and datasets of DB-Hunter to foster reproducibility and facilitate further studies \cite{Codeanddatasets}.
\end{itemize}

\begin{figure}[!t] %H为当前位置，!htb为忽略美学标准，htbp为浮动图形
\centering %图片居中
\includegraphics[width=\linewidth]{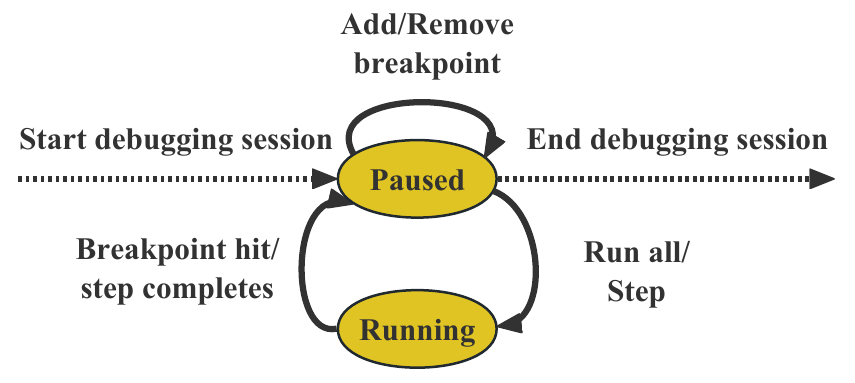} %插入图片，[]中设置图片大小，{}中是图片文件名
\caption{The debugger as a finite state machine} %最终文档中希望显示的图片标题
\label{fig.2} %用于文内引用的标签
\end{figure}

\section{Background and Related Works}
\label{sec:motivaton}
In this section, we present the background of interactive debugger testing and differential testing, along with related work in these areas.
%Before delving into the discussion of our methodology in the following section, we would like to introduce the key concepts behind interactive debugger testing and differential testing, as well as related works.
% , as well as related works.

\subsection{Testing Interactive Debugger}
When engineers debug using a debugger, they set breakpoints at the desired locations and then run the debugger. 
The debugger causes the program to pause execution at the breakpoint, enabling engineers to interactively inspect variable values or identify the location of bugs in the program. 
The debugger receives two types of input: the program to be debugged (for example, Verilog source code) and debug actions. 
Debug actions are commands issued by the user to guide the debugging session, such as setting breakpoints and step-by-step execution of a program. 
In automated testing, it's necessary to generate these debug actions. 
Each sequence of actions must begin with the addition of at least one breakpoint; otherwise, the debugger will never pause during the testing process. 
In addition, the debugger is interactive. 
This means that during a debug session, debug actions are interleaved with the execution of the program being debugged, rather than being issued all at once before starting the debugger.

\figurename~\ref{fig.2} presents a more detailed depiction of the debugger's operations during testing, where the debugger continuously toggles between the paused and running states. Specifically,

\begin{itemize} 
\item[$\bullet$] Debugger program commences in a paused state, whether it's a program yet to start debugging or a program paused during execution. In these states, debugger could capture the current output and issue the next debug action. The program only resumes execution and the debugger switches from running to paused state after executing a debug action (e.g., Run all or Step).
\item[$\bullet$] The debugger remains in the running state until it encounters a breakpoint, completes a step, or finishes an execution task, at which point it switches back to a paused state.
\end{itemize}

\begin{figure}[!t] %H为当前位置，!htb为忽略美学标准，htbp为浮动图形
\centering %图片居中
\includegraphics[width=0.75\linewidth]{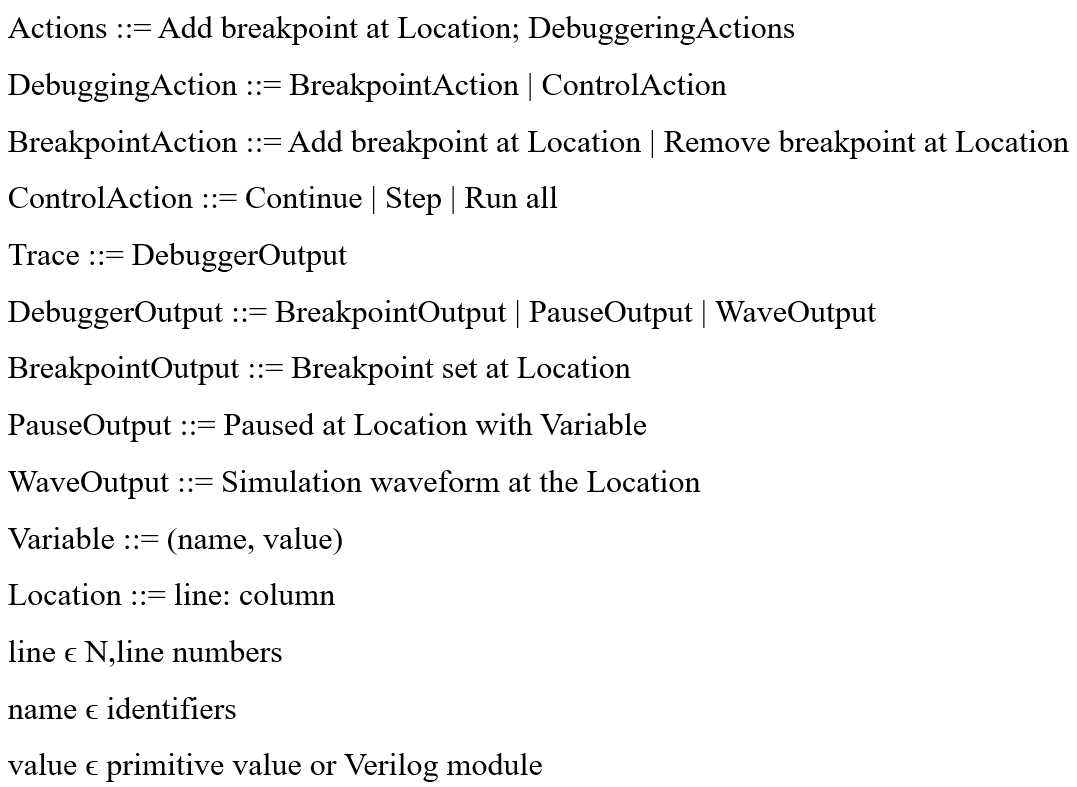} %插入图片，[]中设置图片大小，{}中是图片文件名
\caption{Grammars for debugging actions and traces} %最终文档中希望显示的图片标题
\label{fig.2.1} %用于文内引用的标签
\end{figure}

Finally, the paused state of the debugger records debugging traces. \figurename~\ref{fig.2.1} illustrates the syntax of these debug traces, which encompass three types of debugger outputs:

\begin{itemize}
\item[$\bullet$] Breakpoint output: This part captures the results after setting breakpoints in the debugger.
\item[$\bullet$] Paused state output: This part records the state when the debugger pauses at a breakpoint or upon completion of step execution.
\item[$\bullet$] WaveOutput: This part represents the waveform output in the debugger. DB-Hunter records the state of waveform output at the time of the pause.
\end{itemize}

While debuggers are crucial as widely used development tools, automated testing for debuggers has only recently gained attention.
Lehmann et al. \cite{lehmann2018feedback} employed a differential testing \cite{mckeeman1998differential} approach for the automated testing of JavaScript debuggers. They introduced a feedback-driven testing generator approach capable of producing a series of length-configurable debug actions for the initial test run. 
% Lehmann et al. \cite{lehmann2018feedback} introduced a feedback-driven test generator approach to test debuggers by differential testing \cite{mckeeman1998differential}.
Due to the interactive nature of debuggers, the possibility of pre-generate all test inputs is eliminated. 
This is akin to prior work on interactive GUIs \cite{ermuth2016monkey,hu2011automating,pomeranz2023sharing,machiry2013dynodroid,guo2021hybrid,banerjee2013graphical} automation testing. 
However, testing them due to the interactivity of debuggers presents issues similar to testing at the UI level. 
Although there are existing UI-level test generators, such as those for web applications \cite{artzi2011framework,ermuth2016monkey,uchihara2012h} or for Android \cite{hu2011automating,machiry2013dynodroid,kumar2020methodology,kong2018automated}, they are not easily adaptable to debugger testing. 
The reason being that UI-level events, like traversing code, are closely linked to debugging code, a facet not considered in existing UI-level testing techniques. 
Recently, Sandro Tolksdorf \cite{tolksdorf2019interactive} and others tested the widely used JavaScript debugger in the Chromium browser through mutation, introducing new possibilities for debugger testing.

However, the debugger in JavaScript is typically used for debugging within web browsers. It offers features such as breakpoint setting, variable monitoring, and call stack tracing to assist developers in identifying and resolving code issues. In contrast, the debugger in FPGA development tools is primarily used for hardware description languages (such as Verilog) and related tools, providing functionalities pertinent to FPGA design, such as signal tracing, waveform analysis, and timing constraint adjustments, to help engineers better understand hardware design behavior and performance. Due to the differing targets and functionalities of JavaScript and Vivado debuggers, the methodologies proposed by Lehmann et al. \cite{lehmann2018feedback} for testing JavaScript debuggers are not applicable to testing FPGA chip debugger tool debuggers. However, our DB-Hunter is perfectly suited for FPGA chip debugger tool debuggers, effectively detecting bugs.

\subsection{Differential testing and debugging traces}

The initial RTL design \texttt{Verilog} undergoes mutation to generate equivalent mutant inputs, \texttt{Verilog\textsubscript{A}}, \texttt{Verilog\textsubscript{B}}, ... \texttt{Verilog\textsubscript{N}}. Subsequently, these mutant inputs are processed by executing the FPGA Debugger, resulting in \texttt{Trace\textsubscript{A}}, \texttt{Trace\textsubscript{B}}, ... \texttt{Trace\textsubscript{N}}. In the end, compare these outputs, and if it is discovered that one output differs from the others, it indicates that the input corresponding to this output triggered a bug \cite{utting2023differential,chowdhury2020slemi}.

For simple equivalent transformations, such as the one depicted in \figurename~\ref{fig.1}(a), where the relationship between \texttt{assign X} and \texttt{assign Y} (bitwise operations equivalent mutation, i.e., double negation) is equivalent, their outputs should also be equivalent. In other words, the debugger's behavior on subsequent inputs after transformation should be identical to its behavior on initial inputs. However, for more complex equivalent input transformations, such as when modifying the program being debugged, output relationships need to consider the changes introduced by input transformations. One example of such changes is the reference location used by the debugger. If input transformations alter the underlying program by inserting or deleting lines (as shown in \figurename~\ref{fig.1}(b) where a blank line is inserted in the second row), it is expected that these changes will be reflected in the debugger's output. Specifically, the reported positions will be offset by the number of inserted or deleted lines. These differences must be taken into account when comparing the outputs between initial and subsequent runs.

In recent years, differential testing has found increased application across various domains, particularly in the realm of simulation testing, such as symbolic execution engines \cite{kapus2017automatic}, binary lifters \cite{kim2017testing}, x86 disassemblers \cite{paleari2010n} and Simulink testing \cite{guo2022detecting,jiang2023partition,aranda2020algorithmic,li2023simulink}. 
This trend has offered new insights for employing differential testing to test debuggers in FPGA chip debugger tools. 
Shafiul et al. \cite{chowdhury2020slemi} introduced SLEMI, which focused on the application of mutation detection in a specific domain. 
SLEMI tests MATLAB simulation tool Simulink by trimming unexecuted code and inferring specified data types. 
Guo et al. \cite{guo2022detecting} presented COMBAT, a approach utilizing controllable dummy blocks to construct mutation testing relations, addressing the single-variant generation issue in SLEMI \cite{chowdhury2020slemi}.

\begin{figure}[!t] %H为当前位置，!htb为忽略美学标准，htbp为浮动图形
\centering %图片居中
\includegraphics[width=1\linewidth]{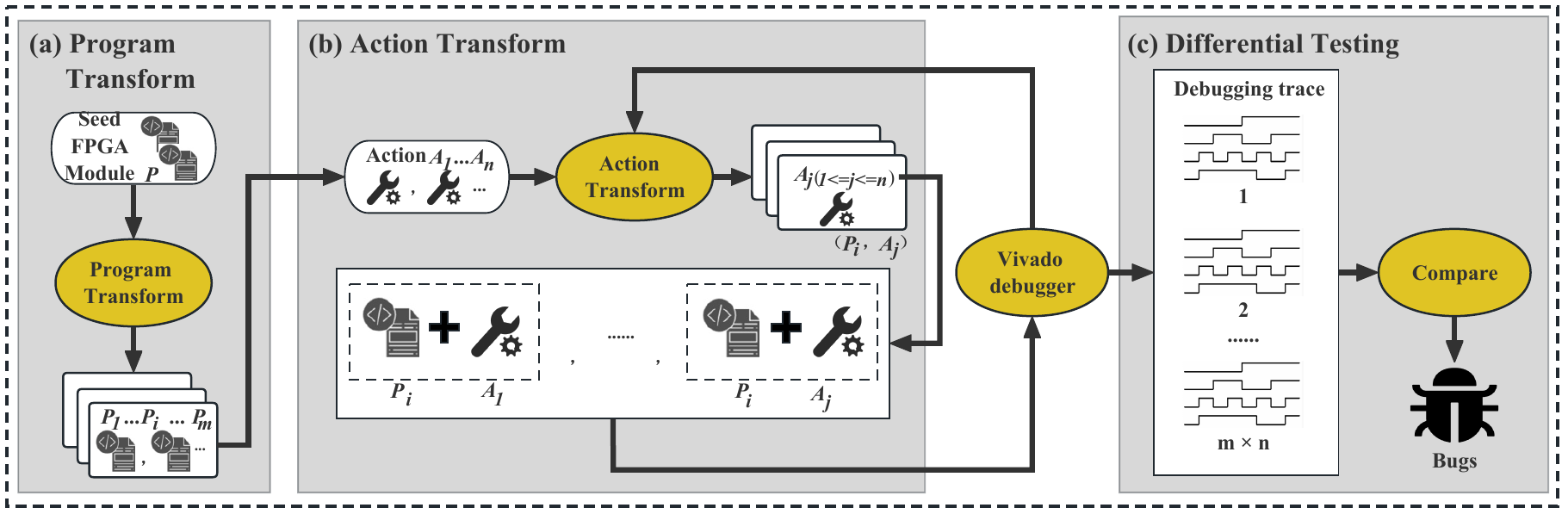} %插入图片，[]中设置图片大小，{}中是图片文件名
\caption{The overview of DB-Hunter} %最终文档中希望显示的图片标题
\label{fig.4} %用于文内引用的标签
\end{figure}

However, existing mutation testing approaches cannot be directly applied to the testing of debuggers in FPGA chip debugger tools. 
This is because the debugger necessitates active user intervention. 
Developers need to manually initiate the debugger, set breakpoints, or monitor expressions before executing the program for debugging purposes. 
Therefore, when testing a debugger, the interactive nature of the debugger must be considered.

\subsection{EMI testing}

Equivalence Modulo Inputs (EMI) is a form of differential testing initially applied in compiler testing. EMI ensures equivalence by generating variant programs that produce equivalent outputs under a given test input. There are three main mutation methods based on EMI differential testing: Orion \cite{le2014compiler}, Athena \cite{le2015finding}, and Hermes \cite{sun2016finding}. Orion \cite{le2014compiler} generates variant programs by selectively removing unexecuted statements from dead code regions. Athena~\cite{le2015finding} focuses on inserting code into these dead code regions. In contrast, Hermes \cite{sun2016finding} is notable for mutating both live and dead code regions to generate equivalent RTL design variants. Once these RTL design variants are generated, both the original RTL design and its RTL design variants undergo differential testing in the compiler. Discrepancies in outputs between the original and variant programs indicate potential issues within the compiler.

\section{DB-Hunter approach}
\label{sec:approach}

In this section, we provide a detailed description of DB-Hunter. An overview of DB-Hunter is presented in Section III.A. Sections III.B - III.D specifically elaborate on the three components of DB-Hunter: the program transformation component, the debug action transformation component, and the differential testing component.

\subsection{Overview}
\label{sec:overview}
We present DB-Hunter, an interactive differential testing approach for detecting bugs in the Vivado debugger.
The framework of DB-Hunter, as depicted in \figurename~\ref{fig.4}.
Firstly, DB-Hunter employs RTL design transformations (RTL design transformations are shown in Table~\ref{tab:1}) on the original RTL designs to generate more complex RTL design variants. 
Subsequently, the RTL design variants are randomly combined with the (debug action transformations are shown in Table~\ref{tab:1}). 
The combined RTL design and debug actions transformations are input into the Vivado debugger, producing a series of debugging traces. 
If iterations are performed, the transformed RTL design and debug actions undergo transformation again. 
Finally, interactive differential testing is employed to compare these debugging traces for consistency. 
Since these RTL design variants are equivalent to the original RTL design, they should yield the same output as the original RTL design. 
If the debugging traces are inconsistent, it demonstrates that DB-Hunter has identified a potential bug.

The implementation of DB-Hunter is primarily outlined by Algorithm \ref{algorithmic.1}. 
Firstly, a original RTL design $s$ is selected and a debug action $a$ is chosen from the set of debug actions.
Subsequently, the original RTL design $s$ undergoes RTL design transformation to produce a mutated RTL design variants $s'$ , and the debug action $a$  undergoes debug action transformation to generate a transformed debug action $a'$. 
The specific process of RTL design transformation is indicated in lines 10-16 of Algorithm \ref{algorithmic.1}, while the specific process of action transformation are outlined in lines 17-23. 
Then, DB-Hunter executes the transformed program and debug action in the debugger, capturing the debugging traces $t$, $t'$ and $t''$ generated during program execution. 
Finally, DB-Hunter employs interactive differential testing to validate these debugging traces. 
If debugging trace generated by the Vivado debugger differ from other debugging traces, it is considered as a potential bug.

\begin{algorithm}[!t]\normalsize
  \caption {DB-Hunter Approach} % 名称
  \label{alg:1}
  \begin{algorithmic}[1]
     \Statex \textbf{Input:}seed modules $\{S\}$, actions $\{A\}$, MAX-iteration $\{M\}$
     \Statex \textbf{Output:} Reported Bug
\Procedure{DB-Hunter}{$S$, $A$, $M$}:
     \For{$s$ $\subset$ $S$, $a$ $\subset$ $A$}      
          \State $t$ $\gets$ Vivado\_Debugger($s$, $a$);
          \State $t^\prime$ $\gets$ Vivado\_Debugger(PRO($s$), $a$);
          \State $t^{\prime^\prime}$ $\gets$ Vivado\_Debugger($s$, ACT($a$));
          \If {$t^\prime$ $\ne$ $t$  \textbf{or} $t^{\prime^\prime}$ $\ne$ $t$ \textbf{or} $t^{\prime^\prime}$ $\ne$ $t^\prime$}
              \State return Report Bug
          \EndIf
     \EndFor
     \Function {PRO}{module $s$} // Program Transformations 
          \State $s^\prime$ $\gets$transform(s, ChooseOperator\_Pro())
          \If {$M$ \textgreater 1}
              $M$ = $M$ - 1;
              \State return PRO($s^\prime$)\
          \EndIf
          \State return $s^\prime$\
     \EndFunction
     \Function {ACT}{action $a$} // Action Transformations 
          \State $a^\prime$ $\gets$transform(a, ChooseOperator\_Act())
          \If {$M$ \textgreater 1}
              $M$ = $M$ - 1;
              \State return ACT($a^\prime$)\
          \EndIf
          \State return $a^\prime$\
     \EndFunction
\EndProcedure
  \end{algorithmic}
\label{algorithmic.1}
\end{algorithm}

% Table '转换方式介绍'
\begin{table}[!t]\normalsize
% \footnotesize
  \centering
  \caption{Transformations for Debuggers.}
  % \caption{TRANSFORMATIONS FOR DEBUGGERS.}  
    \begin{tabular}{lp{25em}}
    \toprule
    \textbf{Transformations} & \textbf{Description} \\
    \midrule
    \multicolumn{2}{l}{\textbf{RTL Design Transformations}} \\
      Assignment conversion & When there is no data interaction, blocking assignment and non-blocking assignment can be converted to each other. \\
    Integer literal to expression & Replacing \textit{2’h3} with \textit{2’h1 + 2’h2} has no effect. \\
    Bit mutation & Replacing \textit{wire1} with \textit{$\sim$($\sim$(wire1))} has no effect. \\
    Remove unreachable loops & Removing unreachable code has no impact. \\
    Deleting unnecessary reference files & Removing unreachable reference files has no impact. \\
    \multicolumn{2}{l}{\textbf{Debug Action Transformations:}} \\
    Add breakpoint & Adding a breakpoint at line \textit{l} has no effect other than pausing at line \textit{l}. \\
    Breakpoint sliding  & Setting breakpoint at line \textit{l} and sliding to $l’$ should be equal to directly setting it at $l’$. \\
    If-else statement testing  & Any debug actions will not be executed on branches that are not executed. \\
    Step for loop & Whether to enter the for loop and whether the number of for loops is accurate. \\
    Code Folding & Folding irrelevant code and determine whether the folded breakpoints exist. \\
    \bottomrule
    \end{tabular}%
  \label{tab:1}%
\end{table}% 

% The overall input for DB-Hunter includes the Vivado debugger under test and the program and debug actions used for the initial test run. Our RTL designs are randomly generated using Verismith \cite{herklotz2020finding}(the RTL designr Fuzzer). Since there are only three debug actions in Vivado: breakpoint, run all, and step, we use a randomly generated method to generate a series of breakpoints, and then run all or step and check whether the correct location paused. The general output of DB-Hunter is the comparison results of debugging traces. If there are differences, DB-Hunter will report a potential bug. Lastly, each warning undergoes manual inspection. If new bugs are detected, they are reported to Xilinx officials.

\subsection{RTL Design Transformation Component}
\label{sec:}
In this section, we mainly introduce the program transform component of DB-Hunter. 
In order to make RTL designs trigger bugs more easily, we use program conversion to enhance the complexity of RTL designs. 
The type of transformation focuses on modifying the source code of the program (our program transformations take place within the design files) for debugging purposes, with alterations in debugging actions merely to align with the modified program. 
The specific process of program transformation is indicated in lines 10-16 of Algorithm \ref{algorithmic.1}. 
DB-Hunter proposes five program transformation approaches based on the characteristics of the Verilog language: transformation of non-blocking assignment to blocking assignment, equivalent mutation of bit operations, replacing equality operators with expressions, deleting unreachable loop statements, and deleting unnecessary reference files.
After program transformation, DB-Hunter obtains more complex and diverse RTL designs and provide them to the Vivado debugger to obtain more comprehensive test. 

\begin{figure}[!t]
	\centering
	\subfigure[Before conversion]{
		\begin{minipage}[b]{0.49\textwidth}
			\includegraphics[width=1\textwidth]{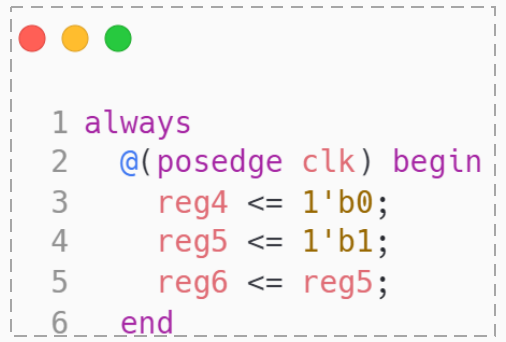}
		\end{minipage}
              \hspace{-4mm}
		\label{fig.5}
	}
    	\subfigure[After conversion]{
    		\begin{minipage}[b]{0.49\textwidth}
   		 	\includegraphics[width=1\textwidth]{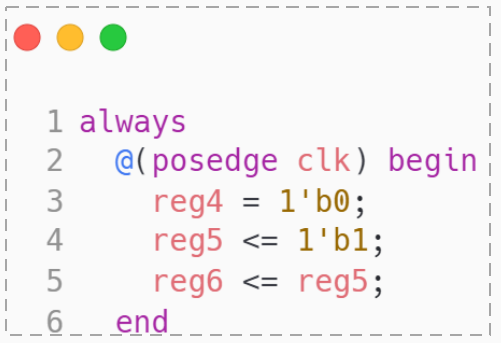}
    		\end{minipage}
		\label{fig.5}
    	}
	\caption{An example of assignment conversion}
	\label{fig.5}
\end{figure}

\subsubsection{\bfseries Assignment Conversion}

For a certain register, when there is no data interaction within the same begin-end block, both blocking and non-blocking assignments to this register have no impact on its final stored data.

As shown in \figurename~\ref{fig.5}(a), non-blocking assignments are used (equivalent to concurrent execution). When the clock signal clk rises, the values of \texttt{reg4}, \texttt{reg5}, and \texttt{reg6} are changed at the end of the ``always" block, and the values used for assignment are those stored before the previous rising edge of the clock. In the end, the value of \texttt{reg4} in line 3 is ($1'b0$), the value of \texttt{reg5} in line 4 is ($1'b1$), and the value of \texttt{reg6} in line 5 is the value stored in \texttt{reg5} before. In contrast, with blocking assignments (equivalent to sequential execution), when the clock signal clk rises, the values of \texttt{reg4}, \texttt{reg5}, and \texttt{reg6} immediately change. Ultimately, the value of \texttt{reg4} immediately becomes the value of ($1'b0$), the value of \texttt{reg5} is ($1'b1$), and the value of \texttt{reg6} becomes the value stored in \texttt{reg5} at that moment, which is ($1'b1$).
Due to \texttt{reg4} in line 4 does not involve data interaction, and therefore blocking assignments and non-blocking assignments will not affect the final result stored in register \texttt{reg4}. In this situation, the two assignment approaches are equivalent and can be interchanged. The final result of our conversion is shown in \figurename~\ref{fig.5}(b).

\begin{figure}[!t] %H为当前位置，!htb为忽略美学标准，htbp为浮动图形
\centering %图片居中
\includegraphics[width=0.95\linewidth]{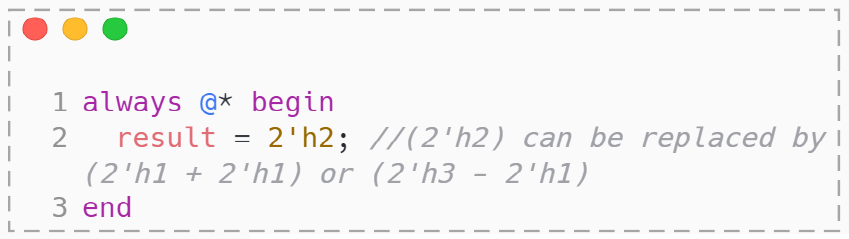} %插入图片，[]中设置图片大小，{}中是图片文件名
\caption{Example of replacing the identity operator with an expression} %最终文档中希望显示的图片标题
\label{fig.pro_bit} %用于文内引用的标签
\end{figure}

\subsubsection{\bfseries Integer Literal to Expression}
For any integer `$n$', DB-Hunter can replace the integer `$n$' with an addition or subtraction expression (e.g., $n = n - 1 + 1$) without modifying the value of the integer `n'. 
\figurename~\ref{fig.pro_bit} shows an example of replacing the identity operator with an expression in Verilog. 2'h2 in line 2 can be replaced with ($2'h1 + 2'h1$) or ($2'h3 - 2'h1$).

% For example, (2'h2) can be replaced by (2'h1 + 2'h1).

% {\verb|Addition|}: ($n - 1 + 1$), for example, ($2'h2$) can c

% {\verb|Subtraction|}: ($n + 1 - 1$), for example, ($2'h2$) can be replaced by ($2'h3 - 2'h1$)

During the replacement process, we need to pay attention to the fact that bit consistency must be maintained. The two-bit value must be replaced with the two-bit value, otherwise unnecessary bit bugs will be introduced.

\subsubsection{\bfseries Bit Mutation}
Given that Verilog stores waveform data in binary form after simulation, performing two bitwise negations (e.g., flipping each bit from 0 to 1 and from 1 to 0) yields a result identical to the original binary value. Using this property, we perform double negations on signals during the signal assignment process.

As shown in line 3 in \figurename~\ref{fig.1}(a), Verilog language demonstrates a double bitwise negation on a signal and stores the final result in the output signal `$b$'. We introduce a similar double negation operation within the Verilog design file, which apparently has no effect on the compilation and simulation of Vivado.

\begin{figure}[!t] %H为当前位置，!htb为忽略美学标准，htbp为浮动图形
\centering %图片居中
\includegraphics[width=0.95\linewidth]{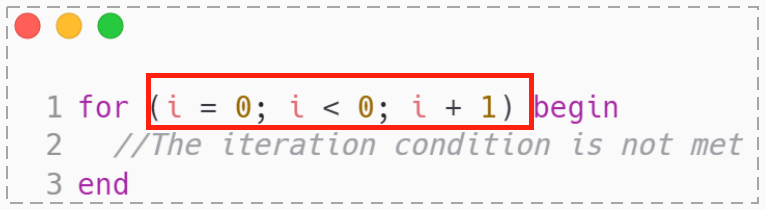} %插入图片，[]中设置图片大小，{}中是图片文件名
\caption{Example of Unreachable Conditions in for Loops} %最终文档中希望显示的图片标题
\label{fig.7} %用于文内引用的标签
\end{figure}

\subsubsection{\bfseries Remove Unreachable Loops.}
In certain loop statements, when the iteration condition is not satisfied, the program does not enter these loop statements. In such cases, the program after removing these unreachable loop statements is equivalent to the original program and does not impact the final result of the program. 

An example is provided in \figurename~\ref{fig.7}. In this case, the initial value of the iteration condition $i$ in the for loop does not satisfy entry into the loop, so the for loop will not execute.

\subsubsection{\bfseries Deleting Unnecessary Reference Files}
During the development process, developers may introduce certain header files to utilize their functionalities. However, as the code undergoes iteration and modification, these header files may become redundant, yet developers forget to remove the corresponding "include" statements, leading to the inclusion of unnecessary header files. When these imported header files have complex dependencies or consume substantial hardware resources, such as registers and logic gates, they can slow down the simulation and synthesis processes of FPGA projects. To facilitate easier maintenance and deployment of FPGA projects and to accelerate simulation and synthesis, we have removed some unused header files. The Verilog programs, before and after the removal of unused header files, are functionally equivalent; therefore, eliminating these unused header files does not result in any change in the simulation outcomes of the RTL design. The approach of DB-Hunter is to initially introduce several header files that will not be used in the RTL design and subsequently remove them. By comparing the simulation results of the two scenarios, if any inconsistencies are found, it indicates that a potential bug has been discovered.

% \begin{figure}[!t]
%   \centering
%   \includegraphics[width=0.95\linewidth]{DB_Hunter_fig/1/figure3.pdf}
%   \begin{center}
%       (a) Principle of Differential Testing
%   \end{center}
%   \includegraphics[ width=0.95\linewidth]{DB_Hunter_fig/1/figure_trace_z.pdf}
%   \begin{center}
%       (b) Debuggering trace
%   \end{center}
%   \caption{Differential Testing}\label{fig.3}
% \end{figure}

\subsection{Debug Action Transformation Component}
\label{sec:Specific code filling}
These transformations exclusively involve debug actions and never alter the mapping between the program and the debugger. In other words, in this scenario, the alterations to the program are akin to an identity change. This implies that, despite potential changes to debug actions, the fundamental correspondence between the program and the debugger remains unchanged. Such an identity change ensures that the behavior of the program after the transformation is equivalent to that of the original program, without introducing significant alterations.
The specific process of action transformation is indicated in lines 17-23 of Algorithm \ref{algorithmic.1}. 
Given that Vivado's debugging actions are limited to breakpoints, run all, and step, our debugging action transformation approaches for Vivado, tailored to its specific circumstances and the characteristics of the Verilog language, involve the following five types: adding breakpoints, breakpoint sliding, breakpoint testing in if-else statements, breakpoint testing in for loop, and code folding.
% There are currently four types: adding breakpoints, breakpoint sliding, breakpoint testing in if-else statements, breakpoint testing in for loop.

\subsubsection{\bfseries Adding Breakpoints.} Since the debugger is interactive, we can manually add breakpoints anywhere. We expect breakpoints to be independent of each other, which means that adding a breakpoint to a program that already has multiple breakpoints set should not modify the existing breakpoints (e.g., disable or move them).
DB-Hunter inserts a new breakpoint ``add\_bp $l_{new}$" at a random location during the initial test run. This also requires runtime information from the Vivado debugger to function correctly. Subsequently, when compare debugging traces between test runs, we examine all pauses at $l_{new}$ in the subsequent traces. If the requested breakpoint location changes due to breakpoint sliding (e.g., when a breakpoint is added on an empty line), the comparison needs to exclude pauses at the actual location $l'_{new}$.

\subsubsection{\bfseries Breakpoint Sliding.} When engineers add comments or line breaks to their programs, these lines of comments and empty lines typically do not correspond to actual runtime operations. As a usability feature, the debugger should move breakpoints to the next ``appropriate" location when such lines are encountered, rather than simply ignoring breakpoints at these positions.

As shown in \figurename~\ref{fig.1}(b), which showcases the behavior of breakpoint sliding in a debugger. 
In line 2 of \figurename~\ref{fig.1}(b), there is a comment line. If we directly add a breakpoint here, the Vivado graphical user interface will display "No Breakpoint set". The solution proposed by DB-Hunter is to sequentially move the breakpoint to the next line until a new breakpoint is successfully set. This ensures that each action that causes the breakpoint movement directly aligns with the actual position. This serves the same effect as the breakpoints before the breakpoint sliding occurred.

\subsubsection{\bfseries If-else Statements Testing.}
For conditional statements such as if-else, there might be a branch that will never be executed. Even if a breakpoint is placed in this never-executed branch, the program will not pause there, and step-by-step debugging will not lead to the execution of this portion of the code.

\begin{figure}[!t] %H为当前位置，!htb为忽略美学标准，htbp为浮动图形
\centering %图片居中
\includegraphics[width=0.95\linewidth]{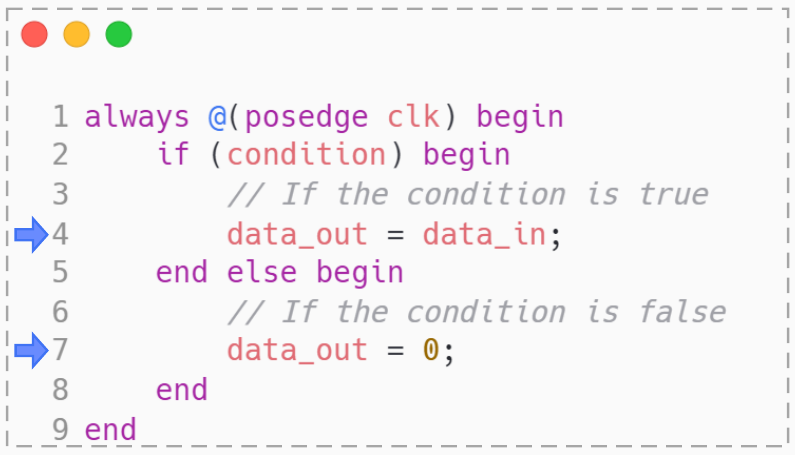} %插入图片，[]中设置图片大小，{}中是图片文件名
\caption{Example of breakpoint testing in if-else} %最终文档中希望显示的图片标题
\label{fig.action_ifelse} %用于文内引用的标签
\end{figure}

\figurename~\ref{fig.action_ifelse} depicts an example of an if-else statement, with breakpoints added at both line 4 and line 7. If the condition is true, the program pauses at line 4; otherwise, it pauses at line 7. If the program pauses elsewhere, it indicates a debugging bug.

% DB-Hunter sets breakpoints in unreachable branches. Subsequently, when running the Verilog program on Vivado, if the program pauses execution at this unreachable branch, it serves as evidence of a bug in the debugger.

\subsubsection{\bfseries Step for loop.}
In the case of a for loop, there are scenarios involving unreachable loops and reachable loops. When the loop is unreachable (as shown in \figurename~\ref{fig.7}), breakpoints within the loop body are not executed, and the program does not pause execution at this location, resulting in zero executions of breakpoints within the loop body. When the loop is reachable, the number of executions of breakpoints within the loop body equals the number of iterations of the loop.

\begin{figure}[!t] %H为当前位置，!htb为忽略美学标准，htbp为浮动图形
\centering %图片居中
\includegraphics[width=0.95\linewidth]{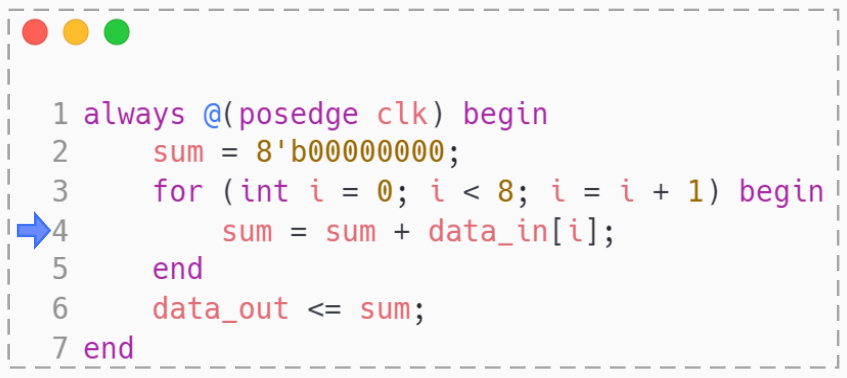} %插入图片，[]中设置图片大小，{}中是图片文件名
\caption{Example of breakpoint testing for reachable for loops} %最终文档中希望显示的图片标题
\label{fig.action_for} %用于文内引用的标签
\end{figure}

\figurename~\ref{fig.action_for} provides an example of a reachable for loop, where it is evident that the for loop iterates 8 times. DB-Hunter sets breakpoints within the loop body (line 4) and records the number of iterations of this loop. If the recorded iteration count differs from the expected count, it can serve as evidence of a debugger bug.

% \begin{figure}[!t]
%   \centering
%   \includegraphics[width=\linewidth, height=0.32\linewidth]{DB_Hunter_fig/1/figure_for.pdf}
%   \begin{center}
%       (a) Equivalent RTL design
%   \end{center}
%   \includegraphics[ width=\linewidth]{DB_Hunter_fig/1/figure_for_wave.pdf}
%   \begin{center}
%       (b) After simulation, comparison of waveforms output by Vivado
%   \end{center}
%   \caption{The process of differential testing to detect bugs in the Vivado debugger triggered by equivalent RTL design.}\label{fig.for}
% \end{figure}

\subsubsection{\bfseries Code Folding.}
During the debugging process, code folding is also a crucial feature. By folding sections of code that do not require immediate attention, developers can more easily focus on the parts of the code they are currently working on, thereby improving code readability. Additionally, folding code can also collapse breakpoints on the corresponding lines, allowing developers to quickly locate critical execution paths during debugging. Since Vivado has a built-in code folding feature, DB-Hunter utilizes Vivado's internal commands (i.e., Tcl commands) to control code folding. After folding the code, DB-Hunter adds breakpoints at appropriate locations. Afterward, DB-Hunter performs debugging operations to verify if the breakpoint's position matches the expectation.

\subsection{Interactive Differential Testing Component}
\label{sec:diff}
DB-Hunter detect bugs in the Vivado debugger through differential testing. 
This approach involves executing different but equivalent RTL designs (such as Verilog programs) in the same environment and comparing their output results. 
It entails examining the behavior generated within the same debugger through two equivalent debugging operations to determine if differences exist. 
Equivalence implies that while the two debugging operations may differ, they still perform the same functionality under the same input and configuration. 
If the outputs differ, it may indicate bugs in the Vivado debugger.

The differential testing in DB-Hunter compares the outputs of the reference model and its RTL design variants under the same input conditions to ensure that modifications do not alter the expected behavior. Formally, let $T$ be the set of RTL designs, $S$ be the Vivado simulation software, and $O(t)$ be the output of system $S$ for RTL design $t$. An equivalent subset of RTL designs 
\( T \subseteq T_{\rm{a}} \) is selected, where $T\textsubscript{a}$ is the complete set of all RTL designs. For all \(t \in T\), the output is calculated as \[O(t) = S(t).\] 

For any two RTL designs 
\({t_1} \in T\) and \({t_2} \in T\), the condition % 
\(O\left( {{t_1}} \right) \ne O\left( {{t_2}} \right)\) is checked, and differences are recorded. If there is at least one inconsistent output: 
\(\exists \left( {{t_1},{t_2}} \right) \in T\) such that 
\[O\left( {{t_1}} \right) \ne O\left( {{t_2}} \right) \Rightarrow exist\;bug,\] a bug is deemed to exist. In this case, DB-Hunter records both the original model S and the variant model V as potential bug cases.

% Specifically, as illustrated in the two examples in \figurename~\ref{fig.5}, the original model and the variant model are equivalent models generated by the program transformation component, meaning these models are functionally equivalent. 
% However, as seen from the waveform graph in \figurename~\ref{fig.9}, after simulation with the Vivado debugger, the waveform graphs generated by the original model and the variant model exhibit significant inconsistencies (occurring between 166ps-167ps and 170ps-171ps). 
% This indicates unforeseen issues with the Vivado debugger during simulation verification, thereby demonstrating the effectiveness of our program transformation component and debugging action transformation component in detecting bugs in the Vivado debugger.

\begin{figure}[!t] %H为当前位置，!htb为忽略美学标准，htbp为浮动图形
\centering %图片居中
\includegraphics[width=0.95\linewidth]{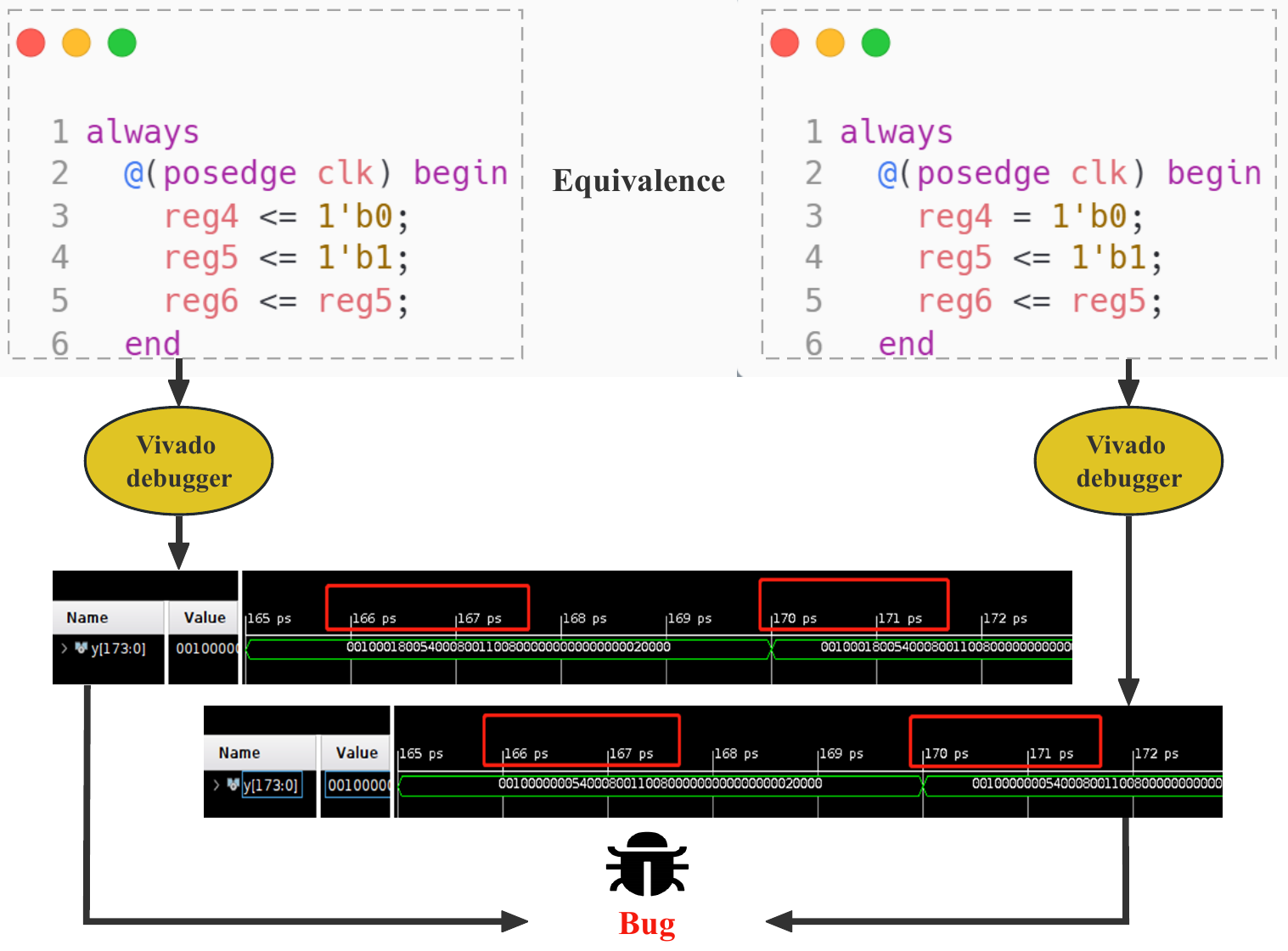} %插入图片，[]中设置图片大小，{}中是图片文件名
\caption{A case study for differential testing after program conversion} %最终文档中希望显示的图片标题
\label{fig.df} %用于文内引用的标签
\end{figure}

\textbf{Case Study.}
For the equivalent variant models generated by the FPGA seed model and DB-Hunter's program transformation component as well as debugging action transformation component, DB-Hunter's differential testing component detects bugs by comparing the output results of the original model and the variant model. 
A case study of bug found by DB-Hunter as illustrated in \figurename~\ref{fig.df}\footnote{\url{https://support.xilinx.com/s/question/0D54U00006pRBdDSAW/vivado\%E4\%B8\%AD125133ps\%E4\%BF\%A1\%E5\%8F\%B7\%E7\%95\%B8\%E5\%8F\%98?language=en_US}}. 
Specifically, the function of this code is to set the value of \texttt{reg4} to 0, the value of \texttt{reg5} to 1, and then transfer the value of \texttt{reg5} to \texttt{reg6} upon each rising edge of the clock signal.
Due to the absence of data interaction within the same begin-end block for \texttt{reg4}, no matter whether non-blocking assignment or blocking assignment is performed on \texttt{reg4}, it will not affect the data stored in it. 
Therefore, we changed the non-blocking assignment of \texttt{reg4} to a blocking assignment.
Since these two code segments are equivalent, running these two equivalent RTL designs in the same Vivado debugger simulation should yield identical output results.

However, as seen from the waveform graph in the lower part of \figurename~\ref{fig.df}, after simulation with the Vivado debugger, the waveform graphs generated by the original model and the variant model exhibit significant inconsistencies (occurring between 166ps-167ps and 170ps-171ps). 
We submitted this bug to Xilinx official and received confirmation from the maintenance personnel (the bug was caused by an issue with the register storing data during the data transmission process in Vivado), Xilinx official indicated that it would be addressed in the next version.

\section{Evaluation}
\label{sec:experiments}

We evaluate the effectiveness of DB-Hunter by applying our implementation to a large set of RTL designs generated by Verismith \cite{herklotz2020finding}. The evaluation focused on the following research questions:

{\bfseries RQ1:} How effective is DB-Hunter in bug detection?

{\bfseries RQ2:} What is the bug detection capability of the various components of DB-Hunter?

{\bfseries RQ3:} What impact does DB-Hunter interactive testing have on discovering bugs?

{\bfseries RQ4:} Which transformation is better at finding inconsistent behavior in DB-Hunter?

{\bfseries RQ5:} What impact does the number of iterations have on bug discovery by DB-Hunter?

\subsection{Experimental setup}
\label{sec:dataset}

DB-Hunter is implemented on Vivado 2022.1 and Vivado 2023.1. The source code and experimental data can be found on GitHub \cite{Codeanddatasets}. DB-Hunter takes the Verilog programs generated by Verismith \cite{herklotz2020finding} (the RTL design Fuzzer) as program input, then provided these seed programs to DB-Hunter for testing. For the setting of the number of lines of code, to be consistent with Verismith’s recommendation, we used seed files between 700 and 1000 lines in size. Our evaluation was conducted on a computer with an Ubuntu 20.04 64-bit system, an Intel Core i9 CPU @ 2.10 GHz, and 120 GB of memory.

To clearly illustrate the root cause of bugs, after DB-Hunter automatically detects a bug, we employed a approach similar to previous work \cite{sun2016finding,tang2021detecting}.
Specifically, we leverage automatic reduction techniques to simplify the RTL design that triggers the bug. This enables Xilinx developers to quickly understand and address the bug. 
In particular, we used the Abstract Syntax Tree (AST) of the bug-triggering Verilog program and executed a binary reduction algorithm. 
If removing RTL design does not eliminate the bug, we discard them from the Verilog program because the trimmed RTL design has no effect on bugs.
Otherwise, removed RTL designs are placed back in their original positions. 
This process continues until no more code can be pruned. 
To avoid reporting duplicate bugs, we manually used failed assertions and back-traces to detect duplication bugs. 
Then, we report the detected bugs as a issue to Xilinx Support\footnote{\url{https://support.xilinx.com/s/topic/0TO2E000000YKY4WAO/simulation-verification?language=en_US}}.
% The Xilinx Support eventually categorizes each issue into new/known/nonbug/pending. 
To make it easier to reproduce our bugs, we published the RTL design that triggered the bugs on GitHub \cite{Codeanddatasets}.

\subsection{RQ1: DB-Hunter detects effectiveness of debugger bugs}
This RQ aims to assess the effectiveness of DB-Hunter in detecting debugger bugs within Vivado. 
The evaluation is conducted by analyzing the underlying causes of inconsistencies detected by DB-Hunter in Vivado. 

Table \ref{tab:addlabel} provides a summary of confirmed bugs along with  types (M: Missed Simulation bug, C: Crash bug), statuses (F: Fixed, N: Not Fixed) and Confirmed (Confirmed: Xilinx officially is confirmed, pending-ver: Xilinx official is verifying). From table ~\ref{tab:addlabel}, we observed that Xilinx has admitted 5 crash bugs and 5 false simulation bugs and 8 simulation bugs are being verified. These crashes are related to different functions, such as the SimpleCheckerConnectionCone function; while simulation bugs are caused by different registers storing data, such as register value transmission bugs when the module is called.

% Table generated by Excel2LaTeX from sheet 'Sheet1'
\begin{table*}[!t]\footnotesize
  \centering
  \tabcolsep=2pt
  \caption{Bugs Reported by DB-Hunter.}
    \begin{tabular}{cllccc}
    \toprule
    \#    & ID    & Title & Type  & State & Confirmed \\
    \midrule
    1     & 6I9QiISAE & Inconsistent output due to changes in file reference sequence & M     & F     & Confirmed \\
    2     & 6nUeKHSA0 & During the synthesis process, Vivado directly closed the program & C     & N     & Confirmed \\
    3     & 6pRBdDSAW & 125-133ps signal distortion in Vivado & M     & N     & Confirmed \\
    4     & 6rVdGjSAK & SimpleCheckerConnectionCone function causes crash & C     & F     & Confirmed \\
    5     & 6uthRDSAY & Bug Detected Due to Top-Level Module Definition Error & C     & F     & Confirmed \\
    6     & 6uupGSSAY & There is a problem with signal waveform change in Vivado & M     & N     & Confirmed \\
    7     & 6yXFXUSA4 & HARTHOptPrepDsp::reduceRegisterSize function causes crash & C     & F     & Confirmed \\
    8     & 7K5T3fSAF & Vivado failure occurs when module instance connects to signal & C     & F     & Confirmed \\
    9     & 7L9pDdSAJ & The breakpoint position can be added does not match the hollow circle & M     & N     & Confirmed \\
    10    & 7QysW8SAJ & An error occurred when debugging after Vivado code folding & M     & F     & Confirmed \\
    11    & 7OfYN0SAN & Simulation results are inconsistent after Vivado deletes useless for loop & M     & -     & pending-verification \\
    12    & 7LAH0TSAX & The 1-bit signal continues to mutate throughout the simulation process & M     & -     & pending-verification \\
    13    & 6qWGAtSAO & Data transmission exception under long bit width conditions & M     & -     & pending-verification \\
    14    & 6tbQSwSAM & A problem with data transmission in module module161 of vivado & M     & -     & pending-verification \\
    15    & 6nVerdSAC & Multiple output data inconsistencies appear after vivado simulation & M     & -     & pending-verification \\
    16    & 7NAx44SAD & The simulation results of vivado and other emulators are inconsistent & M     & -     & pending-verification \\
    17    & 7L8i1USAR & The register data transfer process causes errors in simulation results & M     & -     & pending-verification \\
    18    & 6sbmdnSAA & The results simulated by vivado and other emulators are different & M     & -     & pending-verification \\
    \bottomrule
    \end{tabular}%
    \begin{tablenotes}
    \item Type (M: Missed Simulation bug, C: Crash bug), State (F: Fixed, N: Not Fixed) and Confirmed (Confirmed: Xilinx officially has confirmed, pending-verification: Xilinx official is verifying).
  \end{tablenotes}
  \label{tab:addlabel}%
\end{table*}%

\subsubsection{\bfseries Bug due to breakpoint sliding}
This issue has been documented in problem report \#9. 
\figurename~\ref{fig.1}(a) shows the breakpoint inserted at line 3.
Then, an empty line was added after line 1 (shown on \figurename~\ref{fig.1}(b)). 
Generally, adding an empty line should not substantially alter our program. 
However, during practical execution, we observed that the lines available for adding breakpoints in Vivado didn't shift accordingly when we inserted empty lines or comments. 
This led to discrepancies between the expected and the actual breakpoint positions when adding subsequent breakpoints.

\subsubsection{\bfseries Adding breakpoints at incorrect locations}
This issue has been documented in the problem report \#10. 
To gain a better overview of the code structure, the module section of the design file is typically folded. However, in the folded code, the actual placement of added breakpoints differs from our expectations. 
As shown in \figurename~\ref{fig.8}, we expected to add a breakpoint at line 17; however, Vivado added it at line 8 (the position before code folding). 
This indicates that our code folding is not truly effective, leading engineers to misunderstand their code structure.
This bug has been acknowledged by Xilinx officials and has been rectified in Vivado 2022.1 and Vivado 2023.1.

\begin{figure}[!t]
	\centering
	\subfigure[Expected breakpoint location]{
		\begin{minipage}[b]{0.49\textwidth}
			\includegraphics[width=1\textwidth]{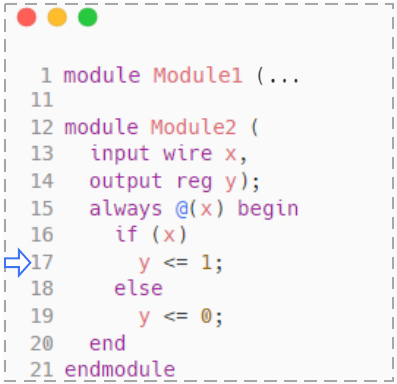}
		\end{minipage}
              \hspace{-4mm}
		\label{fig.8}
	}
    	\subfigure[Actual breakpoint location]{
    		\begin{minipage}[b]{0.49\textwidth}
   		 	\includegraphics[width=1\textwidth]{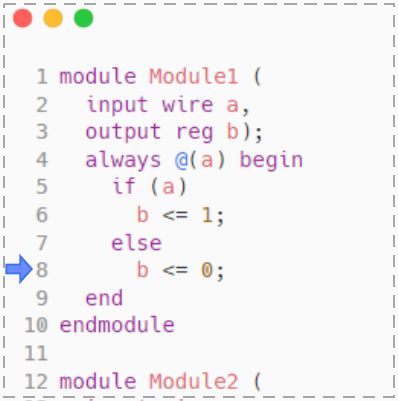}
    		\end{minipage}
		\label{fig.8}
    	}
	\caption{Code folding throws bug}
	\label{fig.8}
\end{figure}

\subsubsection{\bfseries Conversion between blocking assignment and non-blocking assignment causes simulation inconsistency}
This issue has been documented in the problem report \#3.
DB-Hunter modified assignments of registers that did not involve data interaction within the same `begin-end' block, specifically changing non-blocking assignments to blocking assignments. 
DB-Hunter examines registers involved in assignments within the `begin-end' block and sequentially checks if a register exists elsewhere within the block. If it does, indicating a data interaction issue, the corresponding assignment statements cannot be modified to blocking assignments. Otherwise, modifications are made to the assignment statements associated with that register that change non-blocking assignments to blocking assignments. This modification does not impact the final simulation results. If DB-Hunter finds inconsistent simulation results, it proves that this is a potential bug.

\begin{figure*}[!t] %H为当前位置，!htb为忽略美学标准，htbp为浮动图形
\centering %图片居中
\includegraphics[width=0.95\linewidth]{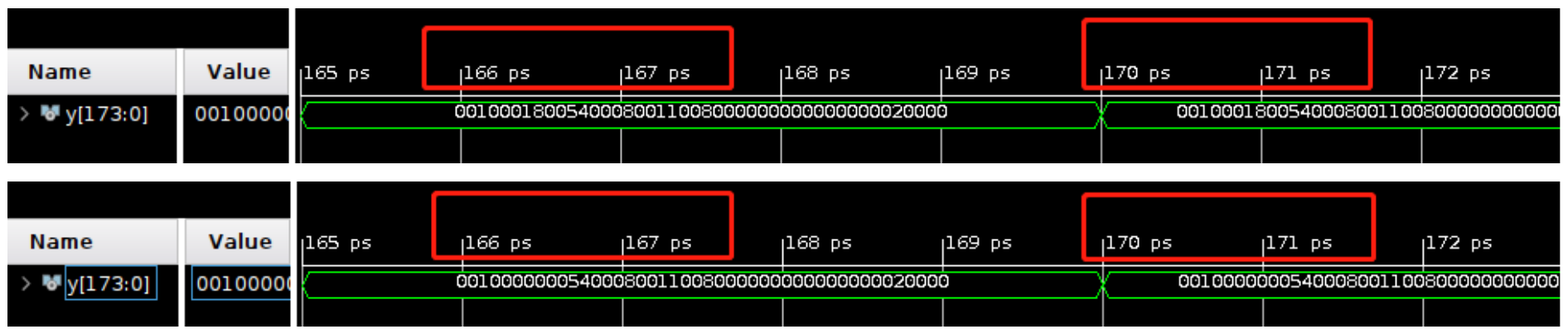} %插入图片，[]中设置图片大小，{}中是图片文件名
\caption{Comparison of simulation results before and after the conversion of blocking assignment and non-blocking assignment} %最终文档中希望显示的图片标题
\label{fig.9} %用于文内引用的标签
\end{figure*}

\figurename~\ref{fig.9} illustrates the simulation results before and after the assignment transformation in the complete program from \figurename~\ref{fig.5} (\figurename~\ref{fig.5}(a) shows the simulation results before the assignment transformation, and \figurename~\ref{fig.5}(b) shows the simulation results after the assignment transformation). 
To eliminate the impact of global reset, DB-Hunter's comparison excludes the initial 100ps. 
It is obvious that Vivado shows a clear difference in the simulation results of the two files between 166ps and 167ps and between 170ps and 171ps. 
This issue has been officially acknowledged by Xilinx as a bug.

% {\bfseries Conclusion: }DB-Hunter is capable of detecting various bugs related to the Vivado debugger. Over a span of three months, DB-Hunter confirmed 10 Vivado debugger bugs.

\subsubsection{\bfseries  topDownPass() function does not manage memory correctly, causing Vivado to crash}

\begin{figure*}[!t] %H为当前位置，!htb为忽略美学标准，htbp为浮动图形
\centering %图片居中
\includegraphics[width=0.95\linewidth]{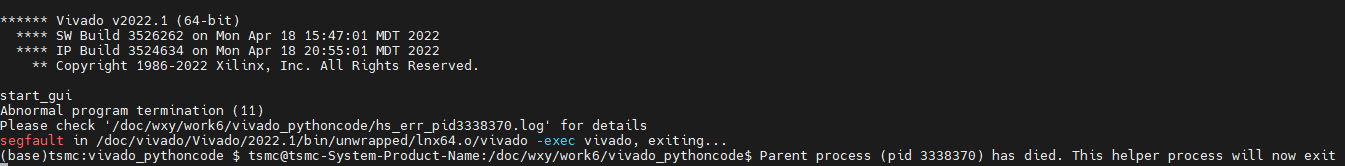} %插入图片，[]中设置图片大小，{}中是图片文件名
\caption{The `topDownPass()` function causes Vivado to crash} %最终文档中希望显示的图片标题
\label{fig_4.2.4} %用于文内引用的标签
\end{figure*}

This issue has been documented in issue report \#5. DB-Hunter successfully identified a problem that caused Vivado to crash. As shown in \figurename~\ref{fig_4.2.4}. Specifically, after applying an equivalence transformation that replaced equality operators with equivalent expressions in the source file, we observed that Vivado crashed during the synthesis of the transformed RTL design. To investigate the cause of the crash, we analyzed the log files generated by Vivado and determined that the issue originated from the internal `topDownPass()` function. This function consumed an excessive amount of memory while processing the RTL design and failed to manage memory properly, resulting in memory overflow or exhaustion.

Upon reporting this discovery to Xilinx, it received significant attention. They responded promptly to the issue by making necessary modifications and optimizations to the `topDownPass()` function. As a result, the latest version of Vivado has addressed this problem, enhancing the software's stability and reliability. This highlights the effectiveness of DB-Hunter in identifying critical software bugs and facilitating proactive improvements in software development processes.

{\bfseries Conclusion: }DB-Hunter is capable of detecting various bugs related to the Vivado debugger. Over a span of three months, DB-Hunter confirmed 10 Vivado debugger bugs. 

% \subsection{RQ2: The impact of components on DB-Hunter's bug detection}

\subsection{RQ2: Compare SOTA (State of the arts) experimental results}

Since we are the first to address issues with the FPGA debugger, this experiment only compares the efficiency of DB-Hunter in detecting bugs by itself.
DB-Hunter involves two types of transformations: program transformation and debug action transformation. To validate the bug detection capabilities of these two transformation types in different scenarios, we applied 2,000 RTL designs separately to randomly select a conversion approach (Random), only program transformation (DB-Hunter(-PT)), only debug action transformation (DB-Hunter(-AT)), or a combination of both (DB-Hunter). In this experiment, we conducted a single iteration on these RTL designs and debugging operations.
% During the testing process, we randomly applied program transformations, debug action transformations, or a combination of both. 
Subsequently, we recorded the results of DB-Hunter testing the Vivado debugger, including instances of inconsistencies and failures.
``Inconsistent" refers to differing results provided by the Vivado debugger across different RTL designs, while ``Failures" indicates abnormal execution or bugs detected by the Vivado debugger.

The experimental results are shown in Table \ref{tab:3}. When DB-Hunter randomly selects a conversion approach, the inconsistency detection rate was 1.60\%. When DB-Hunter specifically performed program conversion, the inconsistency detection rate was 4.06\%. In the case of specially debugging the action transition, DB-Hunter's inconsistency detection rate was only 1.75\%. However, when both transformation types were combined, the inconsistency detection rate of DB-Hunter reached 6.41\%.
It is worth noting that among these 2,000 RTL designs, we excluded some RTL designs that failed due to problems in our code implementation to ensure the accuracy of the test results.
Experimental results show that DB-Hunter is able to detect more inconsistencies when both transformation types are used in combination.

{\bfseries Conclusion: }When program transformations and debug action transformations are used in combination, DB-Hunter is able to detect more inconsistencies.

% RQ2
\begin{table}[!t]\footnotesize
  \centering
  \tabcolsep=3pt
  \caption{Number of Inconsistencies Found in Different Transformation Types}
  % \caption{NUMBER OF INCONSISTENCIES FOUND IN DIFFERENT TRANSFORATION TYPES.}
    \begin{tabular}{lcccc}
    \toprule
    \multicolumn{1}{c}{Transformation types} & \multicolumn{1}{c}{Inconsistent} & \multicolumn{1}{c}{Failures} & \multicolumn{1}{c}{RTL designs} & \multicolumn{1}{c}{Discovery rate} \\
    \midrule
    Random     & 32   & 0    & 2,000   & 1.60\% \\
    DB-Hunter(-PT)     & 81   & 6    & 1,994   & 4.06\% \\
    DB-Hunter(-AT)     & 35   & 0    & 2,000   & 1.75\% \\
    DB-Hunter     & 128   & 3    & 1,997   & 6.41\% \\
    \bottomrule
    \end{tabular}%
  \label{tab:3}%
\end{table}%

% % Table generated by Excel2LaTeX from sheet 'Sheet1'
% \begin{table}[!t]\normalsize
%   \centering
%   \tabcolsep=3pt
%   \caption{Number of Bugs Found by Interactive Testing}
%     \begin{tabular}{cccc}
%     \toprule
%     \makecell[c]{Test \\ case} & Inconsistent & \makecell[c]{Interactive \\ inconsistency} & \makecell[c]{The percentage of bugs \\ found interactively} \\
%     \midrule
%     2,000  & 124   & 57    & 45.97\% \\
%     \bottomrule
%     \end{tabular}%
%   \label{tab:4}%
% \end{table}%

% Table generated by Excel2LaTeX from sheet 'Sheet1'
\begin{table}[!t]\footnotesize
    \centering
    \renewcommand{\multirowsetup}{\centering}
    \tabcolsep=0.2pt
    \caption{Number of Bugs Found by Interactive Testing}
    \begin{tabular}{ccccm{2cm}<{\centering}}
        \toprule
        \multirow{2}{1cm}{RTL design} & \multirow{2}{*}{Inconsistent} & \multicolumn{2}{c}{Interactive inconsistency} & \multirow{2}{2cm}{The percentage of bugs  found interactively}                            \\
        \cmidrule{3-4}             &                               & Adding breakpoints                            & Breakpoint sliding                                                                       \\
        \midrule
        \multirow{2}{*}{2,000}     & \multirow{2}{*}{124}          & 14(24.56\%)                                  & 43(75.44\%)                                                    & \multirow{2}{*}{45.97\%} \\
        \cmidrule{3-4}             &                               & \multicolumn{2}{c}{57}                                                                                                                    \\
        \bottomrule
    \end{tabular}%
    \label{tab:4}%
\end{table}%

% 新添加的实验
\subsection{RQ3: The impact of DB-Hunter interactive testing on bug discovery}
Due to DB-Hunter's interactive nature, it is imperative to understand the impact of DB-Hunter's interactive testing on bug detection. We conducted equivalent transformations on 2,000 RTL designs using a combination of program transformations and debugging action transformations, and subsequently debugged the transformed RTL designs in Vivado. In this experiment, we conducted a single iteration on these RTL designs and debugging operations. As shown in Table \ref{tab:4}, DB-Hunter reported a total of 124 inconsistent warnings. Among these 124 inconsistent warnings, we assessed RTL designs that contained at least one interactive transformation (such as adding breakpoints or breakpoint sliding, as detailed in Table \ref{tab:1}). For example, the bug identified in \figurename~\ref{fig.1} involved a breakpoint sliding operation, illustrating an instance of an interactive transformation. In summary, out of the 124 warnings reported by DB\_Hunter, 57 involved at least one type of interactive transformation, accounting for 45.97\% of the total. Among these, 14 warnings were attributed to issues with adding breakpoints, comprising 24.56\% of the inconsistency warnings, while 43 warnings stemmed from breakpoint sliding issues, accounting for 75.44\% of the inconsistencies identified. This underscores the value of DB-Hunter's interactive differential testing approach. This highlights the value of the interactive nature of our differential testing approach.

{\bfseries Conclusion: }DB-Hunter's interactive testing significantly enhances bug detection. In a study of 2,000 RTL designs, 124 inconsistent warnings were found, with 57 (45.97\%) involving interactive transformations like breakpoint adjustments.

\begin{figure}[!t] %H为当前位置，!htb为忽略美学标准，htbp为浮动图形
\centering %图片居中
\includegraphics[width=0.75\linewidth]{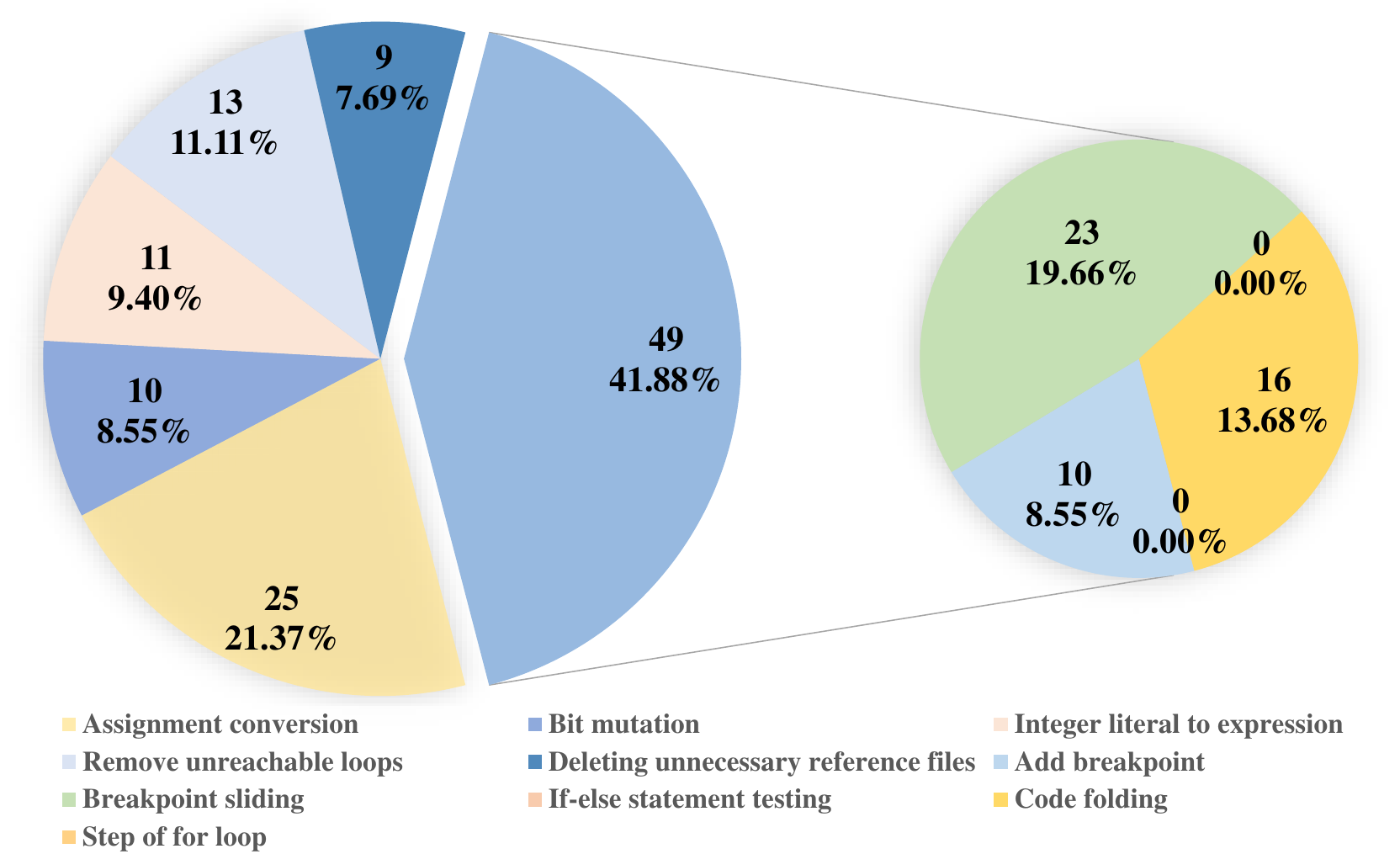} %插入图片，[]中设置图片大小，{}中是图片文件名
\caption{The number of inconsistency warnings found by DB-Hunter
} %最终文档中希望显示的图片标题
\label{fig.rq3} %用于文内引用的标签
\end{figure}

% 新添加的实验
\subsection{RQ4: The most efficient conversion approach for finding inconsistent warnings in DB-Hunter}

DB-Hunter involves 10 types of transformations, each transformation mode has a different ability to detect inconsistent behavior in the Vivado debugger. We applied 2,000 RTL designs to each transformation type separately and conducted a single iteration. Subsequently, we recorded the number of inconsistent RTL designs reported by DB-Hunter for each transformation type in the Vivado debugger. The final result is shown in \figurename~\ref{fig.rq3} that in detecting inconsistent warnings, program transformations account for 58.12\%, while debug action transformations account for 41.88\%.

Among all warnings, the most common inconsistency, with 25 instances, is caused by the transformation from blocking to non-blocking assignments, accounting for 21.37\% of all inconsistent warnings.  This transformation rule is specifically designed for Verilog language features and involves complex timing issues, so it is expected to have the highest number of inconsistencies.

Additionally, 23 instances of inconsistencies, accounting for 19.66\% of all inconsistent warnings, are caused by breakpoint sliding issues, which is the second most common inconsistency.  In all inconsistencies detected in debug action transformations, inconsistencies caused by breakpoint sliding account for more than half of the total.  We attribute this to breakpoint debugging being one of the most common operations, making it prone to issues.  Furthermore, breakpoint sliding issues involve not only graphical interface display but also kernel debugging resources, leading to more resource coordination problems and thus more issues.

The experimental results show that in the subsequent interactive testing of the FPGA debugger, the emphasis of our approach can be placed on the use of blocking and non-blocking assignment transformation and the use of breakpoint sliding strategy to find problems.

{\bfseries Conclusion: }Among all transformation strategies, blocking and non-blocking assignment transformations are the most effective in detecting inconsistent warnings. Meanwhile, Among the four debug action transformations, breakpoint sliding can find more inconsistent warnings.

\begin{figure}[!t] %H为当前位置，!htb为忽略美学标准，htbp为浮动图形
\centering %图片居中
\includegraphics[width=0.75\linewidth]{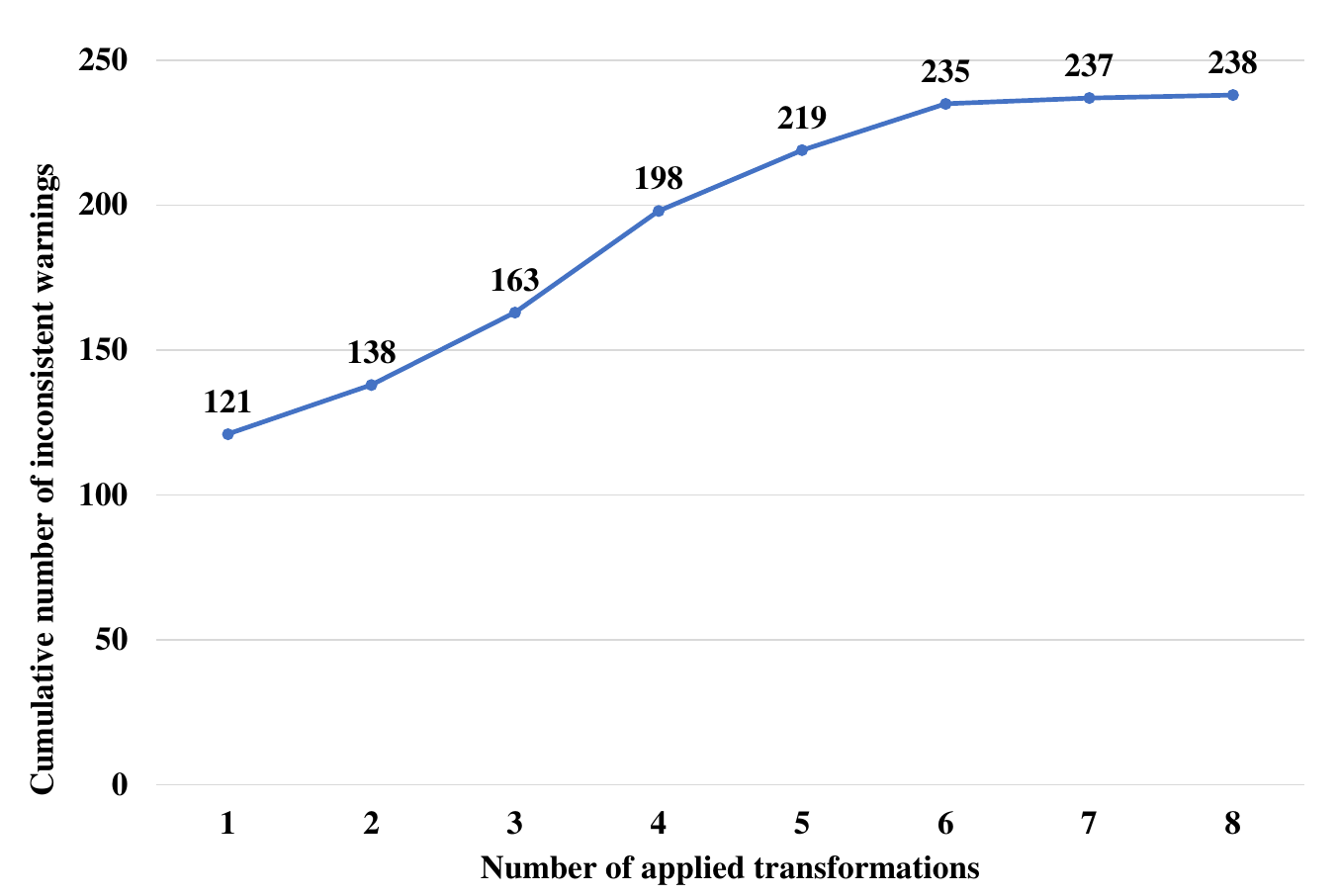} %插入图片，[]中设置图片大小，{}中是图片文件名
\caption{After $N$ iterations, DB-Hunter reports the number of inconsistency warnings} %最终文档中希望显示的图片标题
\label{fig.10} %用于文内引用的标签
\end{figure}

\subsection{RQ5: The impact of the number of DB-Hunter iterations on the number of bugs found}
DB-Hunter is iterative. To evaluate the impact of the number of iterations of DB-Hunter on the number of detected inconsistencies, we ran 8 iterations using 2,000 RTL designs, with a running time of approximately 80 hours. The final results are depicted in \figurename~\ref{fig.10}.
We observed that as the iteration count increased, the number of inconsistent warnings detected by DB-Hunter rose from an initial 121 to 238. However, after reaching 6 iterations, the number of inconsistencies identified by DB-Hunter remained nearly constant. This suggests that DB-Hunter can uncover more issues in the Vivado debugger through multiple iterations, but a higher iteration count does not necessarily yield better results. By the sixth iteration, the RTL designs have covered most of the transformaion scenarios, and the program has become complex enough that subsequent iterations have less impact. Due to the increased time consumption associated with excessive iterations, to ensure optimal performance of DB-Hunter, we have set the maximum iteration count to 6.

Finally, we assessed the runtime efficiency of DB-Hunter. 
Our current implementation takes approximately 30 hours to generate and execute 16,000 RTL designs. 
This timeframe was achieved by concurrently running 8 parallel instances on a single machine (an average of 2,000 RTL designs per parallel instance). 
We believe that the computational resources required for an automated tool of this nature are reasonable and acceptable.

{\bfseries Conclusion: }When the number of iterations is set to 6, DB-Hunter has the best performance.

% \section{Threats}
% \subsection{Internal Threats}
% The accuracy of DB-Hunter and the replication of files from Verismith can affect the effectiveness of our experiments. We conducted a thorough code review to minimize bugs in DB-Hunter. Additionally, we replicated Verismith by referencing the open-source code provided in the original Verismith paper. We have extensively validated their accuracy.

% Although DB-Hunter's program transformations and debugging action transformations can effectively test the debuggers of simulation verification tools, we still aim to further explore potential bugs within these debuggers. Therefore, in future work, we plan to integrate deep reinforcement learning to generate programs and debugging actions that are more likely to trigger bugs, thereby achieving more comprehensive coverage of bugs within simulation verification tool debuggers.

% \subsection{External Threats}
% A key external threat is the possibility of DB-Hunter triggering duplicate defects during the testing process. To mitigate this threat, we have taken measures to remove blocks unrelated to the defects, reducing the variants of defect triggers and enabling a more accurate understanding and analysis of defects. According to verification by Xilinx officials, the bugs we have submitted have not yet reoccurred, indicating the effectiveness of our approach in mitigating this threat.

\section{Conclusion}

In this paper, we present DB-Hunter, an innovative differential testing approach designed for the Vivado built-in debugger. DB-Hunter addresses two primary challenges: establishing equivalence between debug actions and ensuring the consistency of debug actions during program transformations. Given a program requiring debugging and a series of debug actions, DB-Hunter transforms both inputs in a way that introduces changes only in specific aspects while maintaining consistency between the program and debug actions. It then employs differential testing to compare the debugger's outputs. 
%DB-Hunter better accommodates the dynamic nature of debuggers. Unlike traditional differential testing, DB-Hunter introduces the concept of interactive differential testing. This means that we determine transformations not only before the test but also during the execution of the test program, establishing the expected output relationships. DB-Hunter allows for a more comprehensive evaluation of the debugger's performance and stability. 
Experimental results demonstrate that DB-Hunter exhibits exceptional effectiveness and efficiency in testing a widely used real-world debugger. Throughout the testing process, we detected 18 previously unknown bugs in Vivado 2022.1 and Vivado 2023.1, including 10 confirmed as bugs by Xilinx Support, 6 bugs had been fixed.

\section*{Acknowledgment}

This work was supported by the National Natural Science Foundation of China (No.62472062, No.62202079), the Dalian Science and technology Innovation Fund project (No.2024JJ12GX022).

\bibliographystyle{IEEEtran}
\bibliography{mytcad, IEEEabrv}

\end{document}